# Privatização de aeroportos: motivações, regulação e eficiência operacional

Igor Rafael Souza de Brito
Alessandro V. M. Oliveira⇾
Instituto Tecnológico de Aeronáutica, São José dos Campos, Brasil
⇾ Autor correspondente. Instituto Tecnológico de Aeronáutica. Praça Marechal Eduardo Gomes, 50. 12.280-250 - São José dos Campos, SP - Brasil.
E-mail: alessandro@ita.br.

**Resumo**: Neste trabalho vamos abordar alguns assuntos relacionados à privatização de aeroportos na literatura científica, na tentativa de oferecer uma resposta para a seguinte pergunta: a privatização de aeroportos traz resultados positivos? Primeiramente voltamos nossa atenção para as motivações que levam à privatização, considerando as duas principais partes envolvidas, o governo e a iniciativa privada. Afinal, o sucesso desse tipo de decisão será relativo aos motivos que a justificam. Em um segundo momento, vamos considerar a questão regulatória, cuja influência no desempenho financeiro do aeroporto é consolidada na literatura. Na terceira parte, abordamos os principais resultados documentados da privatização, com atenção especial para a eficiência produtiva, cuja melhoria é a mais popular motivação para a privatização.

*Palavras-chave*: transporte aéreo, companhias aéreas, aeroportos.

## I. Introdução

As privatizações de aeroportos têm atraído o interesse da comunidade científica nos últimos anos. Muitos estudos foram publicados abordando, principalmente, os impactos econômicos e sociais atrelados à desestatização desses ativos. Um estabelecimento público, responsável por prestar um serviço essencial à população de uma determinada região, tornar-se propriedade privada e se propor a perseguir objetivos pautados em interesses puramente comerciais já foi motivo de muitos debates acalorados dentro do âmbito da gestão pública. No Brasil, trinta anos depois do início das políticas de privatização mais marcantes, pode-se considerar que o assunto já está, de certo modo, internalizado na sociedade, de sorte que as pessoas já conseguem entender melhor as implicações desse tipo de movimentação. Mesmo assim, ainda é possível encontrar na desestatização controvérsias que alimentam discussões políticas atuais, mesmo que pinceladas, aqui ou ali, com argumentações de cunho majoritariamente ideológico. Por esse motivo, a decisão de se privatizar um aeroporto exige um capital político significativo da parte dos governantes, gerando neles uma justificável preocupação com a opinião pública, de tal sorte que as justificativas para as privatizações apresentadas para a população devem ser cuidadosamente consideradas. Afinal, espera-se que as decisões governamentais tenham por finalidade última beneficiar a população.

Neste estudo, exploramos as dimensões relacionadas à privatização de aeroportos, buscando elucidar se tal processo acarreta em benefícios. Inicialmente, analisamos as motivações por trás da privatização, focando nos interesses tanto do governo quanto do setor privado, visto que o êxito da privatização depende das razões que a fundamentam. Em seguida, examinamos o papel da regulação e seu impacto no desempenho financeiro dos aeroportos, conforme evidenciado na literatura científica. Por fim, investigamos os efeitos documentados da privatização, dando ênfase à eficiência produtiva, frequentemente citada como a principal vantagem da privatização.

## II. As Motivações que Levam às Privatizações

Geralmente considera-se quatro partes interessadas nos efeitos da privatização de aeroportos: o governo, os investidores, as companhias aéreas e os cidadãos, de uma maneira geral. Sejam eles passageiros que fazem uso da estrutura aeroportuária, ou habitantes da cidade ou região onde se localiza o aeroporto que poderiam se beneficiar com a geração de empregos e com o crescimento econômico que normalmente é atrelado ao transporte aéreo. Esses dois últimos players são afetados diretamente pelos acordos firmados entre os dois primeiros, pois são eles os mais importantes clientes dos aeroportos. Nossa discussão, contudo, será focada nos dois principais componentes interessados, a administração pública - que pretende ceder seus bens – e a iniciativa privada, que tem intenção de assumir a operação.

Falando primeiramente sobre o que motivaria a iniciativa privada a querer assumir a operação de um aeroporto, poderíamos resumir bem todo o interesse em um simples "retorno sobre seus investimentos". A possibilidade de adquirir um negócio com características de um monopólio natural – "quando uma única empresa pode oferecer o bem ou serviço para o mercado inteiro a um custo menor do que o fariam duas ou mais empresas" (Mankiw, 2001) - certamente torna a operação aeroportuária muito atrativa para os investidores. As possibilidades de lucro que uma operação com essa característica oferece poderiam justificar os altos valores de ágil que normalmente são atingidos nos leilões das concessões. Contudo, o que é observado em um processo de privatização de aeroportos é que o governo, antes operador daquele aeroporto por meio de uma empresa pública, agora assume o papel de órgão regulador. Os operadores aeroportuários – de qualquer natureza – estão sujeitos a regulações tarifárias, impostas numa tentativa de mitigar o poder de monopólio que um aeroporto normalmente tem por natureza.

A necessidade de se submeter a uma política regulatória pode, de certo modo, arrefecer os objetivos comerciais mais intensos de um operador privado. Contudo, a exploração dessa condição excepcional de poder de monopólio não é o único motivo apontado pelos trabalhos científicos que atrai investidores privados para o setor aeroportuário. De acordo com estudos da Universidade de Waterloo (Mantin, 2012), no Canadá, o que atrai investidores para o setor é a maturidade que a indústria do transporte aéreo vem adquirindo ao longo dos anos. Segundo a publicação, o setor aeroportuário já demonstrou que pode ser autossuficiente economicamente, e que tem alto potencial de geração de receita, principalmente com atividades comerciais não-aeronáuticas que ainda não são exploradas em sua totalidade. Em países em desenvolvimento, onde o crescimento do transporte aéreo se dá, historicamente, de maneira mais tardia, essa característica é ainda mais acentuada hoje em dia. As oportunidades de geração de receita – e de impostos – são ainda pujantes do que em mercados mais consolidados. Ao considerar esse raciocínio, fica claro que o setor privado enxerga nesse tipo de oportunidade o potencial comercial de um aeroporto, que até então é considerado pouco explorado.

Dentro de uma nova maneira de enxergar as atividades comerciais de um aeroporto, esse potencial econômico pode ser identificado ao se considerar essa infraestrutura como uma "plataforma bilateral." Estudos da Universidade de British Columbia, no Canadá, e do Instituto Tecnológico de Aeronáutica descrevem operações aeroportuárias dessa maneira, em que um dos "lados" se concentra a atividade aeronáutica, e o outro é composto pelas atividades comerciais não-aeronáuticas. E a internalização dos efeitos conjuntos que cada lado tem gera valor para o aeroporto. Por exemplo, empresas aéreas se beneficiam com o aumento de passageiros, enquanto os passageiros se beneficiam se houver mais empresas aéreas, mais opções de destinos, um número maior de voos. Uma vez que o "lado aeronáutico" se encontra nessa situação de demanda crescente, a parte comercial do aeroporto se beneficia com o aumento da venda de produtos e serviços ofertados para os passageiros, e para as empresas aéreas.

As taxas de embarque, de pouso e permanência de aeronaves, são exemplos de fontes de receita aeronáuticas, enquanto o aluguel das lojas de varejo, praças de alimentação, aluguel de carro, e outras desse tipo, são exemplos do "lado" não-aeronáutico. Um estudo da Universidade de Sejong, Coréia do Sul, elaborado por Kim & Shin (2001), aponta que à medida que os aeroportos assumem objetivos mais comerciais eles tendem a buscar a maximização da receita não-aeronáutica, dentro da lógica desse novo paradigma voltado ao lucro (também como uma maneira de compensar restrições da política regulatória das tarifa aeronáuticas). Essa dualidade comercial, se bem administrada, pode representar uma significativa fonte de receita para o operador, principalmente se a administração conseguir enxergar oportunidades inovadores de explorar essa condição. Oportunidades favoráveis a essa dinâmica serão criadas caso as lideranças da empresa administradora trabalhem de maneira inteligente com seus dois principais clientes, os passageiros e as companhias aéreas. Afinal, a demanda por transporte aéreo será um benefício aproveitado tanto por empresas aéreas quanto pelo aeroporto, sempre tendo o passageiro como ponto central.

Abordamos, portanto, como um dos componentes interessados na privatização é atraído pelo potencial econômico dos aeroportos modernos. A literatura aponta, porém, que o outro lado conta com diversas justificativas distintas para a privatização. Vamos explorar os principais e mais plausíveis delas. Discutimos, anteriormente, que classe política se preocupa com as consequências da privatização, temendo repercussão negativa na opinião pública. Uma decisão dessa ordem de grandeza pode resultar em graves danos políticos às figuras públicas envolvidas, caso os resultados não sejam tão benéficos quanto o planejado (como por exemplo o Aeroporto Internacional de Viracopos, cuja concessão será relicitada a pedido da concessionária por conta do desempenho financeiro). Consequentemente, o argumento dispensado à população deve ser razoável, destacando, principalmente, as vantagens que serão percebidas pela população.

Aprofundando mais sobre as justificativas das privatizações, um estudo bastante abrangente realizado pela Universidade de Westminster (Graham, 2011), em Londres, apresentou um levantamento das principais motivações de privatização de aeroportos encontrados nas publicações mais importantes. O que se encontrou é que a justificativa mais adotada é a possibilidade de melhorar a eficiência produtiva dos aeroportos – assunto que conta com uma vasta literatura técnica voltada para o entendimento dos aeroportos que estariam na fronteira produtiva, e outros que se desviam dela -, seguido pela oportunidade de conseguir investimentos necessários para obras de expansão e reformas. Na sequência temos a potencial melhoria ou diversificação da administração, e a melhoria da qualidade do serviço prestado. Continuando, fala-se sobre o levantamento de recursos pelo Estado e, por último, uma menor influência da administração pública na operação.

Dentre esses mais populares, daremos atenção aos principais. O primeiro a ser discutido é o referente à possibilidade de arrecadação do Estado. Uma pesquisa realizada pela Universidade de Lisboa (Cruz e Sarmento, 2017) concluiu, após um estudo de caso sobre o Aeroporto de Lisboa, que a principal motivação é a possibilidade de arrecadar dinheiro de maneira rápida, ao ceder a operação aeroportuária. Os leilões públicos nos quais os aeroportos são licitados geralmente fecham com ágios muito acima do esperado. Por exemplo, o consórcio vencedor do leilão da concessão do Aeroporto Internacional de Guarulhos desembolsou pouco mais de 16 bilhões de reais, em 2012, para ter o controle parcial das operações desse aeroporto. Esse valor representou um ágio de 373%. No mesmo leilão, o Aeroporto de Brasília teve ágio de 674%. Os autores também discutem que o governo também se beneficia da condição monopolística dos aeroportos, por conta da valorização que essa situação ocasiona.

A necessidade de investimentos na infraestrutura aeroportuária é outro motivo relevante de se discutir, principalmente em países em desenvolvimento, em que o acesso à recursos pode ser restrito de alguma maneira. Obras de expansão e reformas são necessárias para que a estrutura suporte o crescimento orgânico da demanda e não limite o crescimento do setor. Em vários países, alguns governos decidiram conceder a gestão dos aeroportos ao setor privado como maneira de garantir que obras necessárias fossem realizadas em tempo hábil. No Brasil, por exemplo, frente ao advento de dois grandes eventos esportivos internacionais – Copa do Mundo de 2014 e Olimpíadas do Rio, em 2016 – o país se viu diante de um gargalo estrutural do setor aeroportuário. Essa limitação estaria presente mesmo se não houvesse os dois eventos internacionais, o próprio crescimento da demanda por transporte aéreo do país já alimentava projeções que apontavam para uma restrição estrutural em um futuro próximo. Juntou-se a essas previsões os dois eventos esportivos e se consolidou uma preocupação genuína. Essa necessidade impulsionou o programa de privatizações de aeroportos do país. Foi uma maneira de garantir que os principais aeroportos do país recebessem os investimentos necessários para realizar reformas e expansões necessárias para suportar o volume de passageiros esperados para os eventos.

Por fim, vamos destacar o objetivo mais popular na literatura: o aumento da eficiência produtiva. Muitos estudos se prontificam a estudar a relação entre privatização e melhoria da eficiência da produtividade nos aeroportos. Essa ligação normalmente é construída a partir do pressuposto de que uma administração orientada ao lucro irá gerenciar riscos e custos de maneira mais eficientes, além de explorar potenciais fontes alternativas de receita. Tais medidas são associadas a uma diminuição do custo operacional para as empresas aéreas e, simultaneamente, com o aumento de opções e qualidade de serviços prestados aos passageiros. Um estudo realizado por Poole (1997) ilustra uma situação que deixa essa hipótese da relação entre eficiência e administração privada em evidência. Nos é apresentado o caso do Terminal Internacional 3, do Aeroporto de Toronto, inaugurado em 1991. A empresa estatal canadense estimou que o desenvolvimento desse terminal custaria em torno de um bilhão de dólares canadenses e levaria sete anos para ser concluído. Após a assinatura de um contrato de concessão junto à uma empresa privada, o projeto levou menos da metade do tempo de execução previsto, e a um custo 30% menor. Casos como esse reforçam a ideia de que, diferentemente das empresas públicas, espera-se que uma empresas privada seja muito produtiva, uma vez que, caso não seja, seu sucesso financeiro e valor de mercado será diretamente prejudicado.

Alguns trabalhos científicos, no entanto – como o da Universidade de New South Wales (Kriesler, 2016) - não corroboram a noção da existência de uma relação de causa e efeito entre uma gestão privada e eficiência produtiva. O trabalho exemplificado conclui que não há evidências que sustentem a existência dessa relação, e ainda defende que aeroportos são estruturas em que o tipo ideal de administração é a pública – principalmente por conta da condição de monopólio.

Sintetizando, o que pudemos observar é que, no contexto da privatização de aeroportos, o setor privado é atraído principalmente pelo potencial econômico dos aeroportos, o que poderia garantir o retorno do investimento se explorado em sua totalidade. Os interesses dos governos, por outro lado, estão principalmente na oportunidade de atrair investimentos para reformas e obras de expansão; na possibilidade de captar recursos; e na possibilidade de melhorar a eficiência produtiva dos aeroportos. E o desenvolvimento das localidades, isso é levado em consideração? Por exemplo, formação de hubs (caso de fortaleza), que geraria potencial de investimentos e empregos.

Uma discussão interessante pode ser levantada após as considerações levantadas anteriormente. É muito curioso, afinal, que a busca por uma melhor eficiência produtiva é a motivação mais citada para que o governo chancele uma privatização. Pode-se raciocinar que, quando um aeroporto que se vê afetado por sua infraestrutura limitada enfrenta uma mudança de administração - de pública para privada -, com a intenção de arrecadar os fundos necessários para as devidas intervenções na estrutura, ele não apenas gera fundos para o governo, mas também o economiza uma considerável quantia de dinheiro que normalmente seria alocado para a manutenção do espaço. Após a transferência da administração, portanto, o governo geralmente assume o papel de uma entidade reguladora, em uma tentativa de mitigar os resultados desvantajosos do potencial poder de monopólio associado ao aeroporto, que agora se encontra sob poder do setor privado. Apesar da existência de um mecanismo secundário para incentivar a boa eficiência das concessionárias via políticas de reajustes tarifários - nas quais boas administrações são recompensadas, anualmente, com o direito de reajustar as tarifas à seus interesses (num modelo price-cap) -, nessa nova função, não é a principal prioridade do poder público garantir que a eficiência produtiva do aeroporto seja aprimorada. Cabe-se questionar, portanto, por que essa é a principal motivação que incentiva a mudança de administração da operação aeroportuária, da pública para a privada? Será que o interesse na eficiência de um bem que, em até certo ponto, não está mais sob sua tutela é tão importante assim?

Naturalmente, é plausível supor uma preocupação com o desenvolvimento regional que um aeroporto impulsiona, e na quantidade de empregos que ele gera para uma região. Porém, em um cenário em que apenas um terço dos aeroportos em todo o mundo são lucrativos - de acordo com um relatório do Conselho Internacional de Aeroportos (ACI, 2017) -, pode-se pensar que é muito importante para as intenções de venda dos governos a ideia de que o potencial econômico dos aeroportos públicos é pouco explorado e que, sob um melhor administração, poderia ser mais eficiente e lucrativo. Como veremos adiante, não há consenso sobre o impacto de uma mudança de gestão na melhoria de eficiência. Portanto, uma explicação razoável do porquê esse argumento é importante é que ele serve como uma medida para instigar o interesse do setor privado e valorizar a estrutura, permitindo aumentos de preços das licitações públicas, garantindo uma maior arrecadação na negociação.

### III. A Regulação das Tarifas

A adoção de um objetivo mais comercial, orientado primariamente para a obtenção de lucro, é talvez a mais natural e esperada das mudanças após o processo de privatização de um aeroporto. Com essa transformação, é natural que seja passado um pente fino nas oportunidades comerciais que a estrutura aeroportuária tem a oferecer, numa constante procura por oportunidades lucrativas. Nesse cenário em que o operador e as empresas aéreas buscam objetivos comerciais podem surgir oportunidades de se firmar acordos entre as partes, com propósito de benefício mútuo. Tal consequência é esperada, e as vezes até necessária. Porém, podem surgir oportunidades em que duas partes detêm grande poder de monopólio, de modo que uma parceria, por mais benéfica que seja para elas, se torne nociva para a operação de outras empresas naquele aeroporto. Portanto, na tentativa de coibir práticas comerciais nocivas ao mercado aeronáutico local, que poderia trazer desvantagens desnecessárias para os passageiros, geralmente são propostas regulações tarifárias. Essas regulações são limitações impostas na precificação dos serviços aeronáuticos (e as vezes nos não-aeronáuticos) que buscam contribuir para um equilíbrio operacional nos aeroportos.

É importante, primeiramente, tomar conhecimento dos principais tipos de regulação existentes e suas diferenças. Geralmente as regulações se diferem acerca dos tipos de serviços que serão regulados e sobre como eles serão regulados. Explorando o primeiro caso, podemos classificá-las como regulações single-till e dual-till. Um aeroporto sob regulação single-till tem suas receitas aeronáuticas e não-aeronáuticas simultaneamente controladas. A abordagem de dual-till, por outro lado, tarifa apenas as receitas aeronáuticas buscando impor restrições apenas nas atividades com potencial monopolista. Alguns estudos definem um modelo híbrido (hybrid-till), em que apenas as atividades aeronáuticas são reguladas, com parte dos lucros não aeronáuticos utilizados para subsidiar a parte aeronáutica. Assim, a administração tem mais flexibilidade para traçar estratégias que lhes sejam mais interessantes em determinado momento. Assim como o dual-till, o modelo híbrido também necessita que o aeroporto opere como uma plataforma bilateral, com os lados bem independentes.

Abordando agora os métodos de regulação, os tipos mais comuns que merecem ser citados são a taxa de retorno (RoR) e o teto de preço (price-cap). A regulação RoR é particularmente desenvolvida para controlar lucros excessivos de empresas monopolísticas. Nesse modelo o retorno obtido sobre o capital é limitado. O método regulatório mais comum é o teto de preço. Nessa forma o aeroporto deve respeitar um teto de preço em seus serviços principais - limitando os valores que ele pode cobrar. Dessa forma existe um incentivo para reduzir custos para aumentar o lucro, enquanto na RoR, geralmente os custos são repassados para os clientes, uma vez que a margem é fixada, de certo modo.

Um estudo conduzido por Gillen (2011), da Universidade de British Columbia, deixou uma rica contribuição para a literatura ao apresentar, entre outras coisas, discussões pertinentes acerca da regulação tarifária dos aeroportos. Um dos assuntos levantados é sobre a ineficiência da combinação price cap e single-till. A inclusão de atividades não-aeronáuticas no escopo regulatório pode ser responsável por resultados pouco eficientes. O autor exemplifica com uma situação de aeroporto congestionado, em que o excesso de passageiros ocasiona uma grande arrecadação de receita por atividades comerciais dentro dos terminais, fazendo com que as taxas aeronáuticas sejam reduzidas – por conta da regulação – em um momento onde seria necessário que elas subissem, uma vez que há excesso de demanda. Essa restrição impede que uma precificação condizente seja adotada, afetando diretamente a eficiência financeira do empreendimento. E, em um cenário oposto, onde há falta de demanda, é esperado que o aeroporto (com orientações comerciais) atue com preços baixos para incentivar demanda. Uma situação em que uma política regulatória desse tipo (price cap/single-till) em nada influenciaria, sendo completamente desnecessária.

A discussão evolui, inclusive, para um debate sobre a real necessidade de se regular uma operação aeroportuária. O autor questiona se os aeroportos realmente têm todo esse poder de monopólio que normalmente lhes atribuem. Ele argumenta que os aeroportos mais relevantes têm sim concorrência. Eles não competem apenas com outros aeroportos geograficamente próximos pelos passageiros de uma região, mas sim os outros aeroportos importantes pela condição de "hub", para fornecer conexões para voos de longa duração, e servir de base operacional das grandes companhias aéreas. Nessa situação, dois aeroportos de países diferentes, por exemplo, podem estar competindo por um mercado em comum, e, estando um deles sob uma política regulatória rígida, o aeroporto rival pode se encontrar em vantagem para ganhar a disputa.

E, mesmo no caso de aeroportos que realmente não enfrentam concorrência, é discutível se sua administração realmente tem incentivos para abusar dos preços. O alto custo fixo, e a dependência direta da movimentação de aeronaves e passageiros desencorajariam quaisquer intenções de se aproveitar da condição monopolística. O uso de seu poder de monopólio seria naturalmente limitado, pois, se exercido de maneira abusiva, comprometeria o desempenho financeiro geral do aeroporto (dentro da lógica de uma plataforma bilateral). Podemos observar, por exemplo, o caso dos aeroportos regionais. Esses são bons exemplares de mercados teoricamente monopolísticos que não são atrativos para a iniciativa privada justamente por não ter, na prática, a condição de monopólio. Normalmente estão fora do radar das empresas aéreas para operar como hub por conta de suas estruturas e/ou posições geográficas, e também não atraem o número suficiente de voos para conferir-lhes certas condições comerciais que despertam o interesse privado – destacando que uma companhia aérea pensa sempre em sua malha operacional quando considera adicionar um destino à sua operação.

O autor conclui que o argumento do poder de monopólio não é suficiente para introduzir uma regulação muito restritiva. Em vez disso, políticas mais brandas – chamadas light-handed – seriam mais adequadas para o mercado moderno.

Contudo, alguns estudos importantes oferecem evidências empíricas que atestam a importância de se adotar uma política de regulação. Os resultados obtidos no estudo empírico publicado por Bel e Fageda (2010), do Instituto Universitário Europeu de Florença, indicam que em aeroportos privatizados a ausência de regulação tem efeito aumentador dos preços das tarifas. Em um estudo afim, Adler e Liebert (2014), pesquisadoras da Universidade Hebraica de Jerusalém e da Universidade de British Columbia, encontraram resultados sugerem que a forma de regulação do price cap/dual-till está positivamente associada ao aumento da eficiência produtiva e da receita aeronáutica. No entanto, é importante destacar que o caso base, que era o modelo não regulamentado, apresentou o segundo melhor resultado. Mas o destaque é que a forma de regulação mais flexível apresentou os melhores resultados.

IV. CONSEQUÊNCIAS DA PRIVATIZAÇÃO DE AEROPORTOS

Mas o que se pode esperar de um aeroporto que se tornou privado? Essa certamente é a pergunta que todos os estudiosos do assunto gostariam de ter a resposta. A literatura nos aponta para algumas consequências mensuráveis, mas ainda não apresenta um consenso sobre o resultado dessa mudança de política sob alguns aspectos importantes. Por exemplo, ao retornarmos à discussão sobre as motivações por trás das privatizações podemos nos perguntar: as privatizações deixaram os aeroportos mais eficientes como esperado?

A verdade é que vários estudos foram conduzidos com o objetivo de aferir causalidade entre tipo de administração e eficiência produtiva, mas seus resultados não convergiram para um consenso. Segundo uma importante autora que escreveu sobre o assunto, Graham (2011) - da Universidade de Westminster -, uma possível explicação para os resultados inconclusivos é que os métodos utilizados talvez não sejam os adequados. Alguns autores, como Bezzerra e Gomes (2016) da Universidade de Coimbra, concordam que as métricas de desempenho aeroportuário geralmente falham em representar a realidade de uma operação comercial de um aeroporto, e na verdade deveriam considerar uma perspectiva mais ampla, na qual os indicadores conversem com as necessidades de cada player importante para a operação.

Trabalhos referenciados, como os apresentados por Oum et al. (2006) e por Adler e Liebert (2014) – ambas associados à Universidade Hebraica de Jerusalém e também à Universidade de British Columbia -, atestam que administrações conjuntas – públicas e privadas – são menos eficientes que as demais, mas que não se encontra diferença significativa de eficiência entre aeroportos completamente públicos ou privados.

Duas outras consequências interessantes da privatização observadas, que vale a pena levar em consideração, são o impacto na demanda e a saúde financeira da empresa estatal que cedeu os aeroportos para a iniciativa privada.

Em um estudo nacional, Rolim et al. (2016), do Instituto Tecnológico de Aeronáutica, desenvolvem um modelo econométrico de demanda de viagens aéreas para aeroportos recém-privatizados. Este artigo concentra-se nos aeroportos brasileiros, mais especificamente nos três primeiros a serem leiloados no início da década passada. Os principais resultados indicam que a privatização está diretamente relacionada a um aumento geral da demanda por transporte aéreo. Os autores atribuem esse efeito às mudanças nas relações comerciais entre aeroportos e companhias aéreas, ocasionadas pela nova visão orientada ao lucro dos operadores aeroportuários. Um fruto dessa nova dinâmica comercial entre as partes, que influenciam diretamente na demanda, é a criação – ou expansão – do serviço de desenvolvimento de rotas, ofertado pelo aeroporto. Esse serviço consiste em o próprio Aeroporto oferecer estudos de mercado que possibilitariam a implementação de novas rotas entre ele e outra localidade, ou então de um aumento da frequência de rotas existentes. É uma maneira do aeroporto indicar que existe demanda a ser atendida e, assim, atrair mais empresas aéreas ou um maior número de voos para sua estrutura, aumentando o fluxo de passageiros em seus terminais. Em um estudo da Universidade de New South Wales - Sydney, (Halpern e Graham, 2015) os autores indicaram que 86% dos aeroportos de operação privada que eles consideraram possuem equipe de desenvolvimento de rotas própria, enquanto menos da metade dos aeroportos públicos ofereciam esse serviço. Lembrando, mais uma vez, que em uma configuração de plataforma bilateral, em que as atividades aeronáuticas e não-aeronáuticas são desenvolvidas separadamente, o aeroporto se beneficia ainda mais com o aumento do volume de passageiros transitando por seus terminais, principalmente por conta de exploração da receita gerada pelas lojas e restaurantes presentes na estrutura.

O estudo desenvolvido por Fernandes e Pacheco (2018), da Universidade do Rio de Janeiro, buscava determinar o impacto no desempenho financeiro de uma empresa estatal de administração aeroportuária após a privatização de alguns de seus aeroportos mais importantes. O foco é, também, nos aeroportos brasileiros. O desempenho dos aeroportos administrados pela estatal Infraero, após as ondas de privatizações no Brasil, é o objeto de estudo do trabalho. Analisando dados referentes aos 60 principais aeroportos administrados pela empresa entre 2009 e 2015, os autores observam uma queda em sua participação de mercado após as privatizações. Durante o mesmo período, a demanda geral por transporte aéreo no país aumentou significativamente. Portanto, enquanto os aeroportos públicos perdiam participação de mercado, a demanda por transporte aéreo subia, o que indica que os aeroportos privados estavam absorvendo toda essa nova movimentação. É concluído, portanto, que os aeroportos da Infraero não conseguiram desempenhar bem nessa nova realidade das privatizações, e que os aeroportos regionais foram os principais responsáveis por essa queda de rendimento da estatal. Segundo os autores era esperado pelo governo, porém, que participação privada na administração de aeroportos brasileiros pudesse, de certo modo, possibilitar uma reorganização da Infraero e, consequentemente, uma melhoria de seu desempenho. Essa melhoria não foi observada.

Brito, Oliveira e Dresner (2021) abordam os impactos das privatizações de aeroportos sobre as companhias aéreas. Desenvolvem modelos empíricos para avaliar como a privatização de aeroportos afeta os preços das passagens aéreas, empregando uma metodologia de diferenças em diferenças juntamente com uma regressão de troca endógena para analisar a seleção de aeroportos privatizados. Os resultados dos autores indicam um aumento de 3–3,5% nas tarifas de voos que envolvem pelo menos um aeroporto privatizado, em comparação com rotas operando entre aeroportos públicos, com a dominância de mercado ampliando esse efeito.

## V. Considerações Finais

Quando se trata de uma privatização, podemos apontar que a iniciativa privada é atraída pelo potencial econômico que é um aeroporto moderno, buscando sempre o retorno sobre seus investimentos. Os governos, por outro lado, se interessam basicamente em conseguir investimentos necessários para financiar obras de reforma e expansão e, por muitas vezes, se interessam na arrecadação do valor da venda. A literatura aponta, porém, que o argumento mais utilizado é referente ao interesse em aumentar a eficiência produtiva dos aeroportos. Vimos, porém, que a literatura não endossa essa alegação, ao mesmo tempo que discutimos como ela é importante para os objetivos dos programas de privatização. Afinal, é importante para que as licitações sejam bem-sucedidas, que exista essa ideia acerca da subexploração do potencial comercial dos aeroportos - causada por uma administração pública engessada e que prioriza outros objetivos -, e que uma operação privada saberia bem como aproveitá-lo.

Também foram discutidos os tipos de regulação tarifária que normalmente são adotados, e como que uma separação das atividades reguladas, atrelados à uma flexibilização do regime tarifário, pode ser benéfico para um Aeroporto. Também foi levantada a possibilidade de que a ausência de regulação poderia ser uma boa opção, uma vez que os aeroportos possivelmente não detêm o poder de monopólio que lhes é atribuído. Resultados empíricos, porém, atestam a necessidade de se regular, mas também apontam para um melhor resultado advindo de regulações mais flexíveis.

Por último, podemos concluir após examinar alguns documentos importantes sobre os possíveis resultados da privatização, que não há consenso sobre se a privatização de fato aumenta a eficiência produtiva. Também podemos indicar uma causalidade de demanda de privatização, exemplificada pela preocupação de expandir - ou criar - o serviço de desenvolvimento de rotas. Também abordamos o resultado da privatização para a empresa do governo administrativo do aeroporto depois que seus principais aeroportos foram privatizados. Esses são tópicos importantes que devemos prestar atenção especial aos novos artigos que os abordam e que os cientistas que investigam assuntos relacionados à privatização devem considerar suas pesquisas.